\documentclass[12pt]{article}
\usepackage[margin=1in]{geometry}
\usepackage{fullpage, graphicx, amssymb, amsmath, amsthm, amsfonts, multirow, enumerate, multicol, calc, thmtools, mathrsfs}
\usepackage{authblk}
\usepackage{listings}
\usepackage{xcolor}
\usepackage{tikz}
\usepackage{fancyhdr}
\usepackage{subcaption}
\usepackage{nicematrix}
\usepackage{amsmath}
\usepackage{amssymb}
\usepackage{mathtools}
\usepackage{bm}
\usepackage{hyperref}
\usepackage{natbib}
\usepackage{setspace}
\usepackage{float}
\usepackage{tabularx}

\title{Economic Disparities and Their Relationship to Destructive Health Behaviors in Five Western U.S. States}

\author{Noah Jackson}
\author{Sergey Lapin}

\affil{Department of Mathematics and Statistics, Washington State University}
\date{June 2026}

\begin{document}
\maketitle
\doublespacing
\section*{Abstract}
In this paper, we look at the relationships that economic variables have with adverse health outcomes in the western counties of Washington, Idaho, Oregon, California, and Nevada, with specific emphasis on how suicide rate relates to such economic variables. Data was first gathered from Census and County Health Rankings for the entire United States (for website use and usefulness for future research), cleaned and regression-imputed, and then various exploratory data analysis methods were used, such as PCA, clustering, correlation gathering, linear fittings, and LASSO. PCA and clustering suggested that counties may group according to broader state-level economic patterns, although political interpretations would require additional electoral data. Correlation Analysis, along with LASSO and linear fittings, showed us the destructive variables that connected the most with economic variables (in terms of $R^2$ and correlation values seen), the economic variables that are most and least important in predicting suicide rate, and the possible relationships that suicide rate has with these economic variables.

\noindent\textbf{Keywords}: Suicide Rate, Poverty Rate, Regression Imputation, Principal Component Analysis, Clustering, Linear Fit, Correlation, LASSO, R-Squared, Variance Inflation Factor, Cross Validation

\section{Introduction}

Exploring relationships between economic variables and health-related destructive behaviors gives us multiple benefits. For one, it can help states, counties, cities, and other populations identify economic conditions associated with adverse health outcomes and guide further investigation. Another reason, though not the last, is that it helps us better understand human and societal behavior. 

We will conduct this study at the county and state levels for this topic. Past papers have related more specific variables with others, such as relating income and suicide rate (\cite{repec:fip:fedfwp:2007-12}). This paper was originally focused on suicide rate as being the main concerning destructive behavioral factor to look at, since it has been of major concern in the past few decades; according to \cite{StoneDeborah2018}, suicide rate increased more than 30\% in 25 states from 1999 to 2016. 

Further research in the data gathering process showed that this project could be expanded to much more than just the suicide rate as a destructive behavior to explore. By doing this, we also make the dataset and study more useful for a wider variety of research. Suicide rate is not the only concerning health variable of recent years, too; according to \cite{spencer2024drug}, age-adjusted drug overdose rate increased from 8.2 to 32.6 deaths per 100,000 from 2002 to 2022. Most of these added health variables we will look at, if not all, are important destructive behaviors that county and state leaders look at regularly.

In this paper, we gather data from the government and university sources, use imputation methods to fill missing entries, derive new data columns/variables from already existing ones, do exploratory data analysis (EDA) as well as a detailed correlation analysis, and create and analyze linear models while examining their features. To see more on data gathering, imputation, and deriving new variables, see the Methods (\ref{Methods}) section. To look into the EDA, correlation analysis, and examination of linear models, look at the Results (\ref{Results}) section. Furthermore, looking at the Discussion (\ref{Discussion}) and Limitations (\ref{sec:limitations}) sections is recommended as well.

\section{Methods}
\label{Methods}

\subsection{Data Collection}

Economic variables were gathered from \cite{uscensus_acs5_2024} ACS 5-year county-level data via an API key. The variables that were gathered include median income, poverty rate, Bachelor's plus percentage, unemployment rate, labor force participation, average household size, median gross rent, total occupied housing units, along with owner-occupied and renter-occupied housing units with $x$ occupants per room (where $x$ can be 1.01-1.5, 1.51-2, or $>$2, for a total of 6 more variables). 

With the variables above, derived variables were made from the already present variables in the dataset. The derived variables include Overcrowding, Labor Deprivation, Equivalized Income, and Rent Index Region:

\begin{table}[H]
    \centering
    \begin{tabular}{|c|c|}
    \hline
     Derived variable & Formula \\ 
     \hline
     \text{ Overcrowding Num.} & $\sum_x \text{Owner Occ. with } x\text{ a room } + \sum_x \text{Rent Occ. with } x \text{ a room}$ \\
     \hline
     \text{Overcrowding} & $(\text{Overcrowding Num.})/(\text{Total Occupied units})$ \\  
     \hline
     \text{Equivalized Income} & $(\text{Median Income})/{\sqrt{\text{Average Household Size}}}$ \\
     \hline
     \text{ Rent Index Region} & 
     $(\text{Median Gross Rent})/{\text{Med}(\text{Median Gross Rent)}}$ \\
     \hline
     \text{Real Equivalized Income} & $(\text{Equivalized Income})/({\text{Rent Index Region}})$ \\
     \hline
     \text{Labor Deprivation} & (\text{Unemployment Rate} + (100 - \text{Labor Force Participation}))/2 \\
     \hline
    \end{tabular}
    \label{tab:formulas-for-derived}
    \caption{Formulas for the derived variables}
\end{table}
The overcrowding numerator can simply be interpreted as the number of homes with $>$1 occupants per room. Overcrowding has been linked to health risks in the past, as the World Health Organization recommended in 2018 for national governments to implement policies to reduce overcrowding to help in lowering household health risks (\cite{WHO2018Housing}). Equivalized Income is used to adjust the income of a household based on the household size, making it more accurate and comparable across households of different sizes (\cite{KARVONEN2021100892}; \cite{OECDIDD2012}). Real Equivalized Income gives another normalization to income for the counties; Rent Index Region gives a proportional estimate to Regional Price Parities (RPPs) for the counties, and thus dividing by it adjusts each county's income once more in accordance with differences in county prices. Rent Index Region is also referred to as the median rent index (MRI) in \cite{burns2024geographic} and is first used in \cite{renwick2011geographic}. According to \cite{repec:bpj:bejeap:v:18:y:2018:i:4:p:17:n:8}, authors have noted again and again the importance of adjusting for regional price variation when studying inequality, tax progressivity, and urban agglomeration premiums. Labor deprivation can be seen as the average between the unemployment rate and labor force nonparticipation. This measure is motivated by \cite{serajuddin2015deprived}, which explains how looking at unemployment alone may give an inaccurate picture of a labor market's well-being, and how labor force participation and unemployment can interact with each other.

Beyond this, health variables were obtained from the \cite{uwphi_chrr_2025_analytic_data} County Health Rankings Website. Variables that were selected include suicide rate, drug overdose rate, injury death rate, homicide rate, motor vehicle crash death rate, firearm fatality rate, alcohol impaired driving deaths percentage, along with percentages of adults who report excessive drinking, insufficient sleep, smoking, physical inactivity, and frequent mental distress. This gives us a variety of adverse health outcomes and behavioral health indicators that we can use in order to find patterns between them and the economic variables already selected. 

After gathering the data via Census and the County Health Rankings, there was missing data for the counties. To make the dataset complete, we decided to use Regression Imputation on the missing entries. How the imputation was implemented was first by replacing the missing entries with the respective column (or variable) median. Following this, for each column that had at least one missing entry before replacement, a linear model was fitted using the other columns as predictors, and then we used each respective linear model to predict such missing entries. We replaced missing entries with the median so that we could use counties that didn't have data for all the columns for the fitting process, and also so that we could always have valid inputs for imputation (or, in other words, prediction) on the entries that we wanted to impute on.  For more details on this process, you can see the website that is linked in the Data Availability Statement (\ref{Data Availability}) section.

Common ways to deal with missing values are to replace them with the mean, median, or mode, but Regression Imputation helps in preserving the relationship between missing values and other variables (\cite{Zhongheng2016}). How this implementation can give inaccurate results and possibly even worse inaccurate results for counties that have a multitude of missing entries to begin with is further discussed in the Limitations (\ref{sec:limitations}) section.

\subsection{Statistical Methods}

After the data gathering, we made choropleths for the suicide rate and the poverty rate to get a first impression of the data. Next, we used Principal Component Analysis (PCA) on the economic variables of the dataset (\cite{hotelling1933analysis}). Using PCA on economic variables has been useful in the past: it was found that in a macroeconomic dataset, the first principal component produced the best inflation forecasts from 1959 to 1997 (\cite{repec:eee:moneco:v:44:y:1999:i:2:p:293-335}; \cite{NBERw32068}). However, in this instance, we use PCA in this paper more for visualization instead of forecasting or prediction. Furthermore, $k$-means clustering (\cite{macqueen1967methods}) was used on the PCA data to identify potential patterns (clusters were again used more for visualization than anything else).

After such exploratory data analysis was done, we looked at the correlation that economic variables had with the destructive behavior variables by doing a correlation matrix and then created a graph out of the selected correlations, while filtering out any correlation that was less than 0.4 in absolute value.

We then fitted a linear regression model for suicide rate (as it was supposed to be the only health variable originally) using economic variables as predictors. After fitting the model, we decided to fit a LASSO model (\cite{Tibshirani1996}) instead after seeing extreme Variance Inflation Factor (VIF) values (\cite{Kutner2005}). In the LASSO model, we selected the $\lambda$ regularization parameter by using cross-validation (CV) (\cite{Hastie+al:2009}). We plotted the $\lambda$-coefficient paths for all predictors, which gives us a sense of which predictors (i.e., the economic variables) are most important in predicting the suicide rate for a county.

\subsection{Software and Packages}

The data collection was done in Python, while the statistical methods were done in R/RStudio. The data cleaning process heavily relied on the packages \texttt{pandas} (\cite{mckinney2010pandas}) and \texttt{numpy} (\cite{harris2020numpy}), while the R packages include \texttt{dplyr} (\cite{dyplyrcite}), \texttt{igraph} (\cite{igraphcite1}; \cite{igraphcite2}; \cite{igraphcite3}), \texttt{corrplot} (\cite{corrplotcite}), \texttt{car} (\cite{carcite}), \texttt{glmnet} (\cite{glmnetcite1}; \cite{glmnetcite2}), \texttt{ggplot2} (\cite{ggplot2cite}), \texttt{sf} (\cite{sfcite1}; \cite{sfcite2}), \texttt{tigris} (\cite{tigriscite}), \texttt{viridis} (\cite{viridiscite}), \texttt{patchwork} (\cite{patchworkcite}), \texttt{magick} (\cite{magickcite}), \texttt{kableExtra} (\cite{kableExtracite}).

\section{Results}
\label{Results}

\begin{figure}[H]
    \centering
    \includegraphics[width=0.9\linewidth]{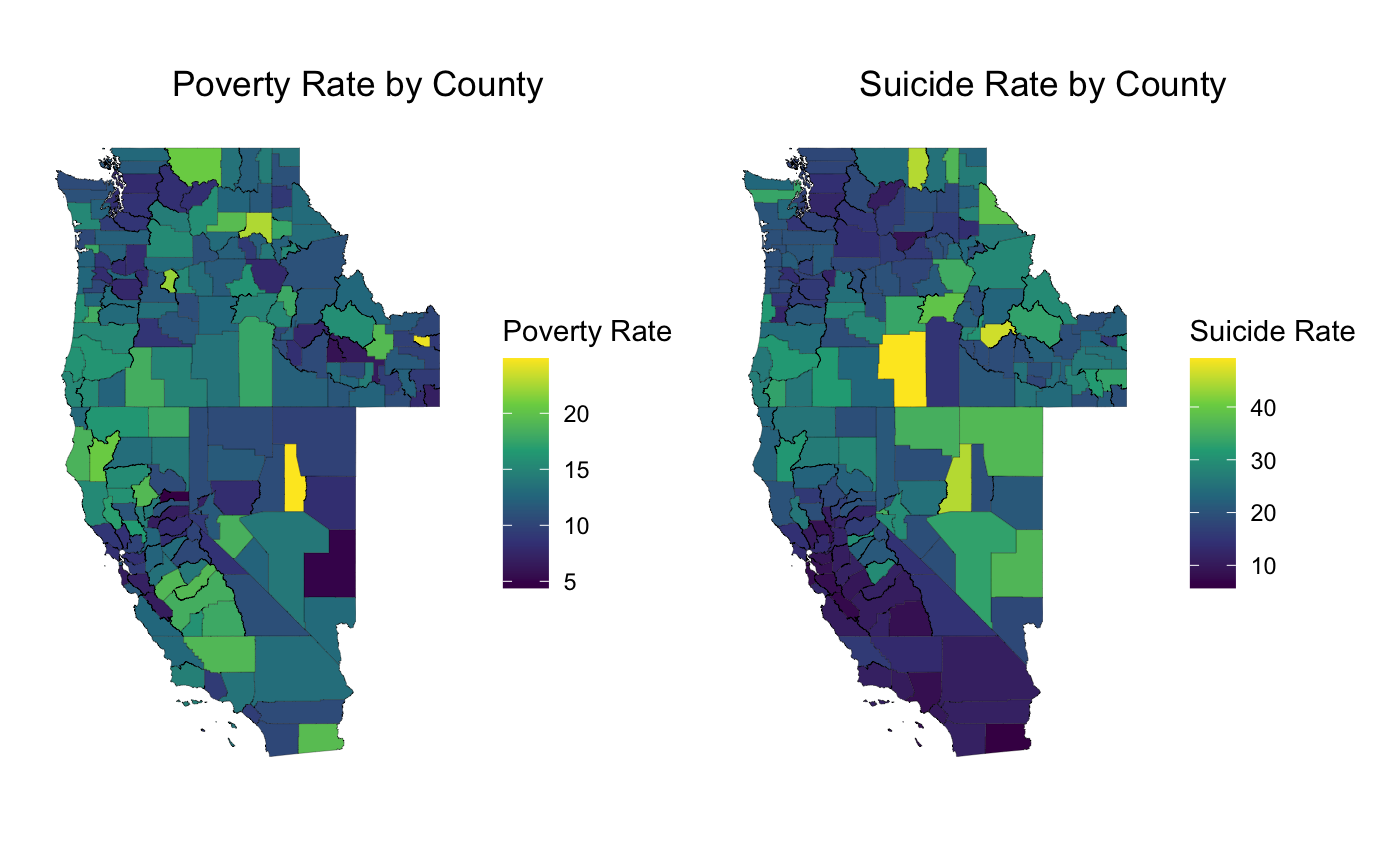}
    \caption{Choropleths for Suicide rate and Poverty rate}
    \label{fig:suicide-poverty-chloropleths}
\end{figure}

Figure \ref{fig:suicide-poverty-chloropleths} gives us a first look at the data by looking at a economic variable and a destructive variable that are popular metrics when talking about this subject of concern. We can see that the poverty rate decreases in counties of larger population (counties close to or around Portland, Seattle, San Francisco, etc.), and is somewhat randomly high in counties that are said to be rural. Suicide rate seems to be quite low in Southern California and Western Washington, while high in more rural counties (though it doesn't seem to be high in the same counties where the poverty rate is high).

\begin{figure}[H]
    \centering
    \includegraphics[width=0.75\linewidth]{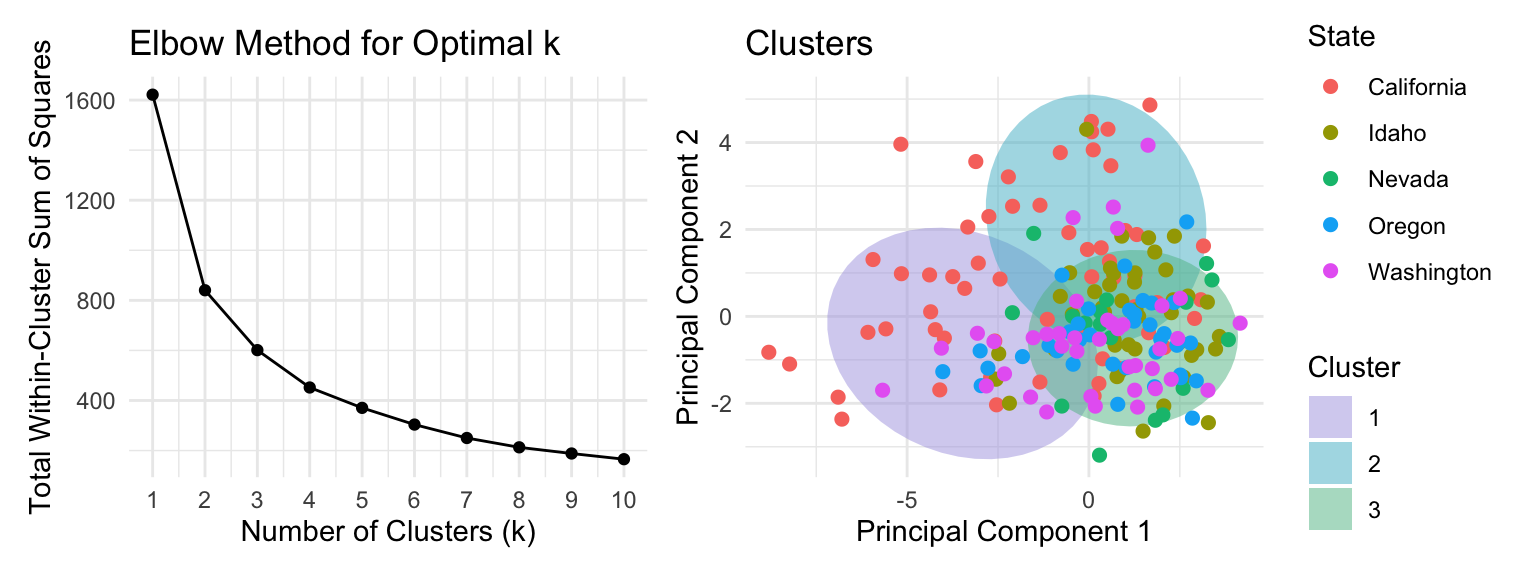}
    \caption{Clustering for the PCA data using the elbow method}
    \label{fig:clustering-elbow-PCA}
\end{figure}

After the choropleths, we explored the use of Principal Component Analysis on the economic variables, giving Figure \ref{fig:PCA-plot-western} below. In this figure, we can see that there are many California counties that tend to be further away from a large portion of the other counties. This makes us ask if we can infer what state a county is in, based only on economic data from the county. We explore this topic by using $k$-means clustering (using the elbow method), as seen in Figure \ref{fig:clustering-elbow-PCA}.
 
\begin{figure}[H]
    \centering
    \includegraphics[width=0.75\linewidth]{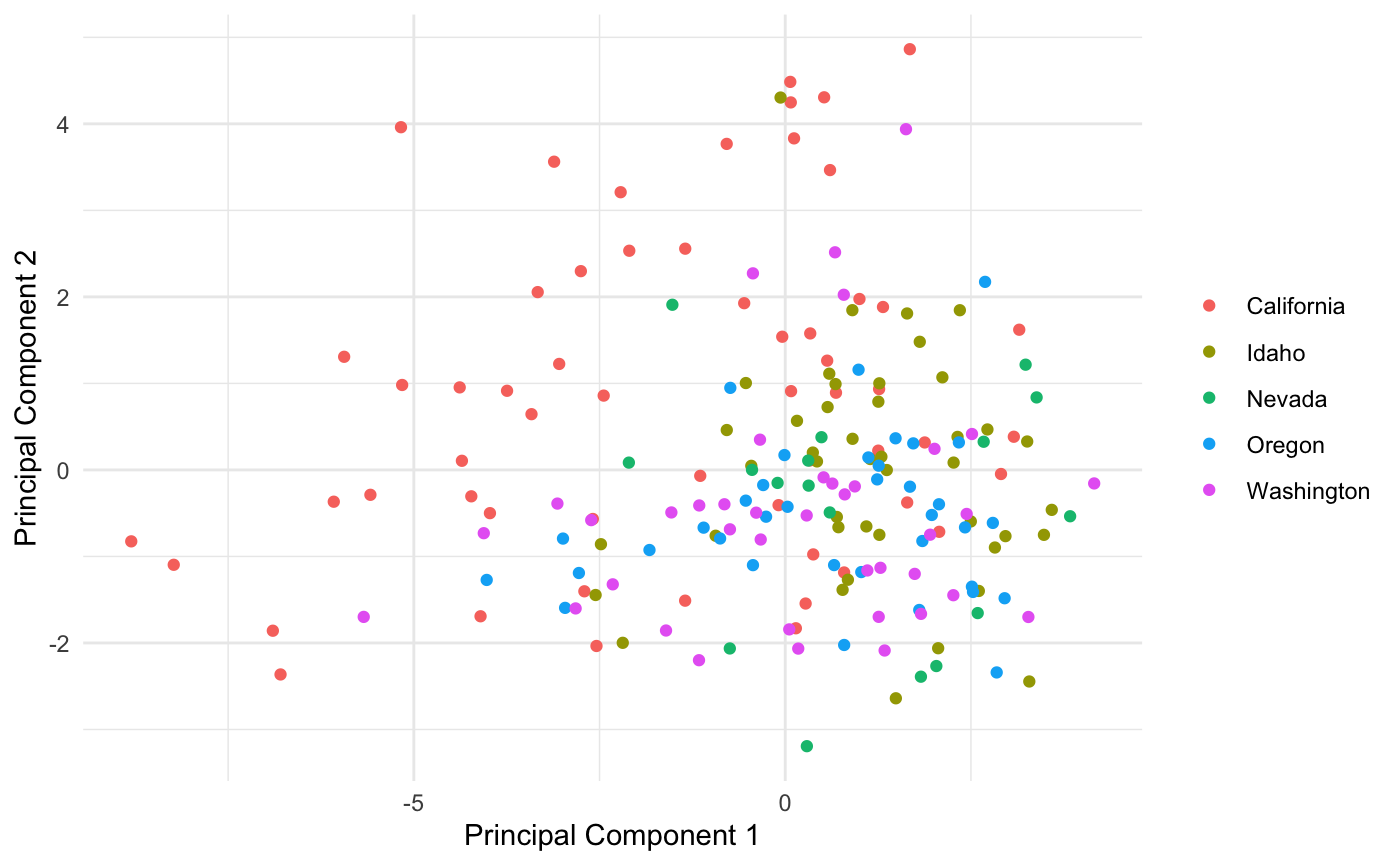}
    \caption{PCA plot done for counties in the five western states via only using economic variables}
    \label{fig:PCA-plot-western}
\end{figure}

The resulting three clusters gotten from the elbow method give us some interesting results. The most notable result is that the third cluster (the green cluster in Figure \ref{fig:clustering-elbow-PCA}) contains almost all the counties from the states of Nevada and Idaho. Secondly, the California counties are split into two clusters, and this result could be due to the larger Euclidean distances that the California counties tend to have between each other in comparison to counties from other states.

We next looked at the correlation between all of our variables in the model, with emphasis on the correlation that the economic variables have with the destructive behaviors.

\begin{figure}[H]
    \centering
    \includegraphics[width=0.9\linewidth]{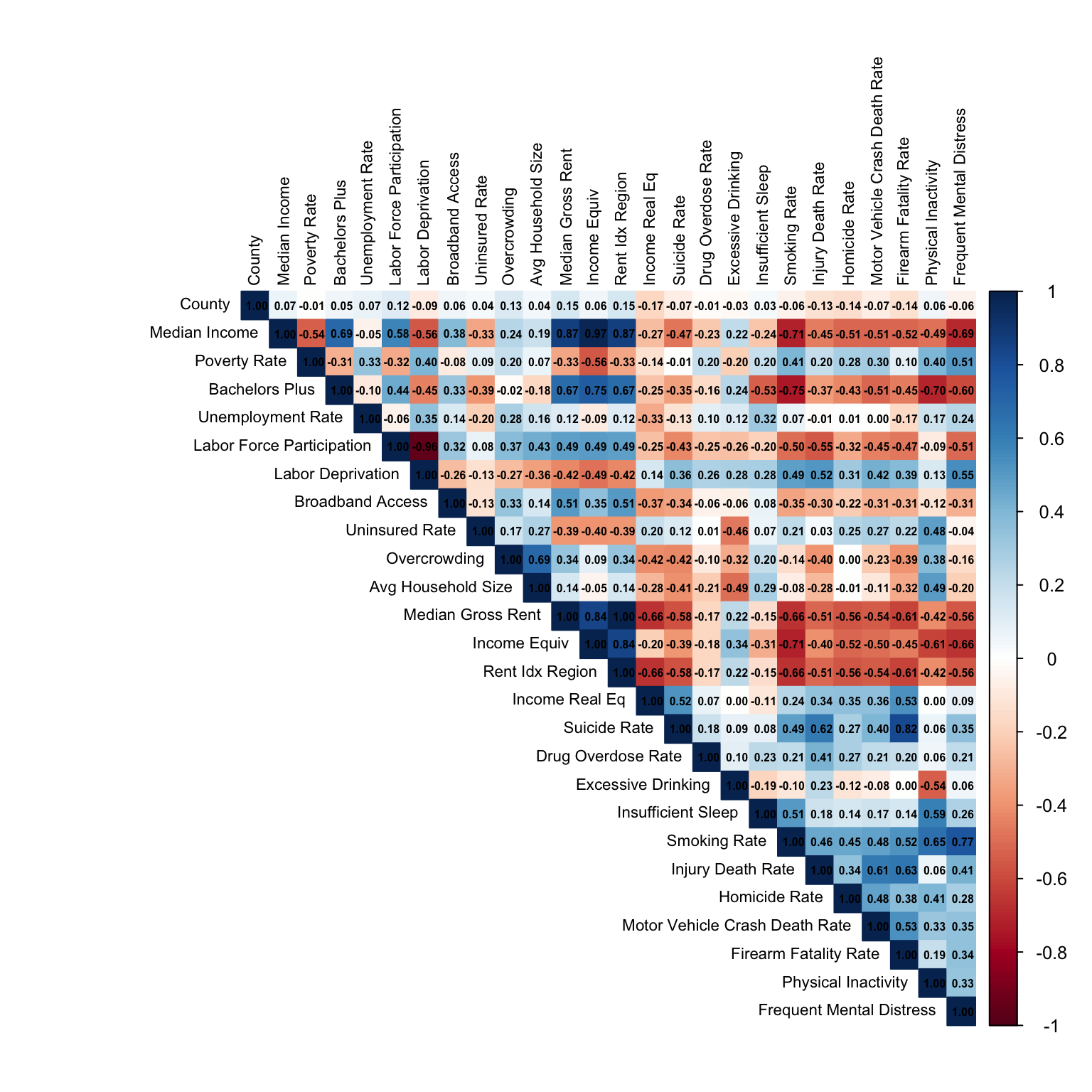}
    \caption{Correlation matrix between all the variables in the data set}
    \label{fig:corr-matrix}
\end{figure}

\begin{figure}[H]
    \centering
    \includegraphics[width=0.9\linewidth]{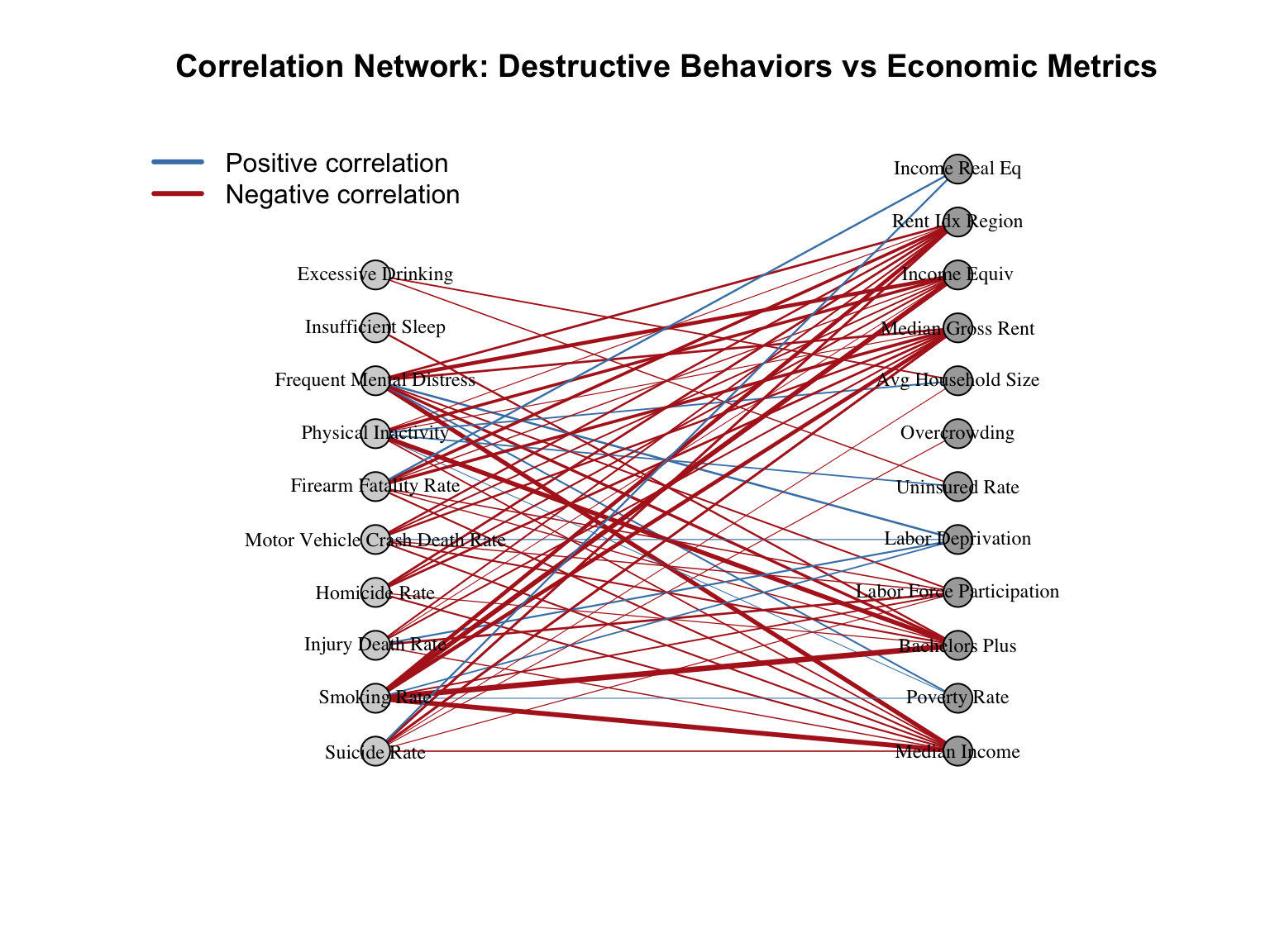}
    \caption{Graph between Destructive variables and Economic variables}
    \label{fig:destr-econ-graph}
\end{figure}
In Figure \ref{fig:corr-matrix}, we see the correlation that many of the economic variables have with each other. We note this as a concern for multicollinearity if we are to make a linear model. We also see strong negative correlations that destructive variables have with economic variables from Figure \ref{fig:corr-matrix} and Figure \ref{fig:destr-econ-graph}.

\begin{figure}[H]
    \centering
    \includegraphics[width=0.7\linewidth]{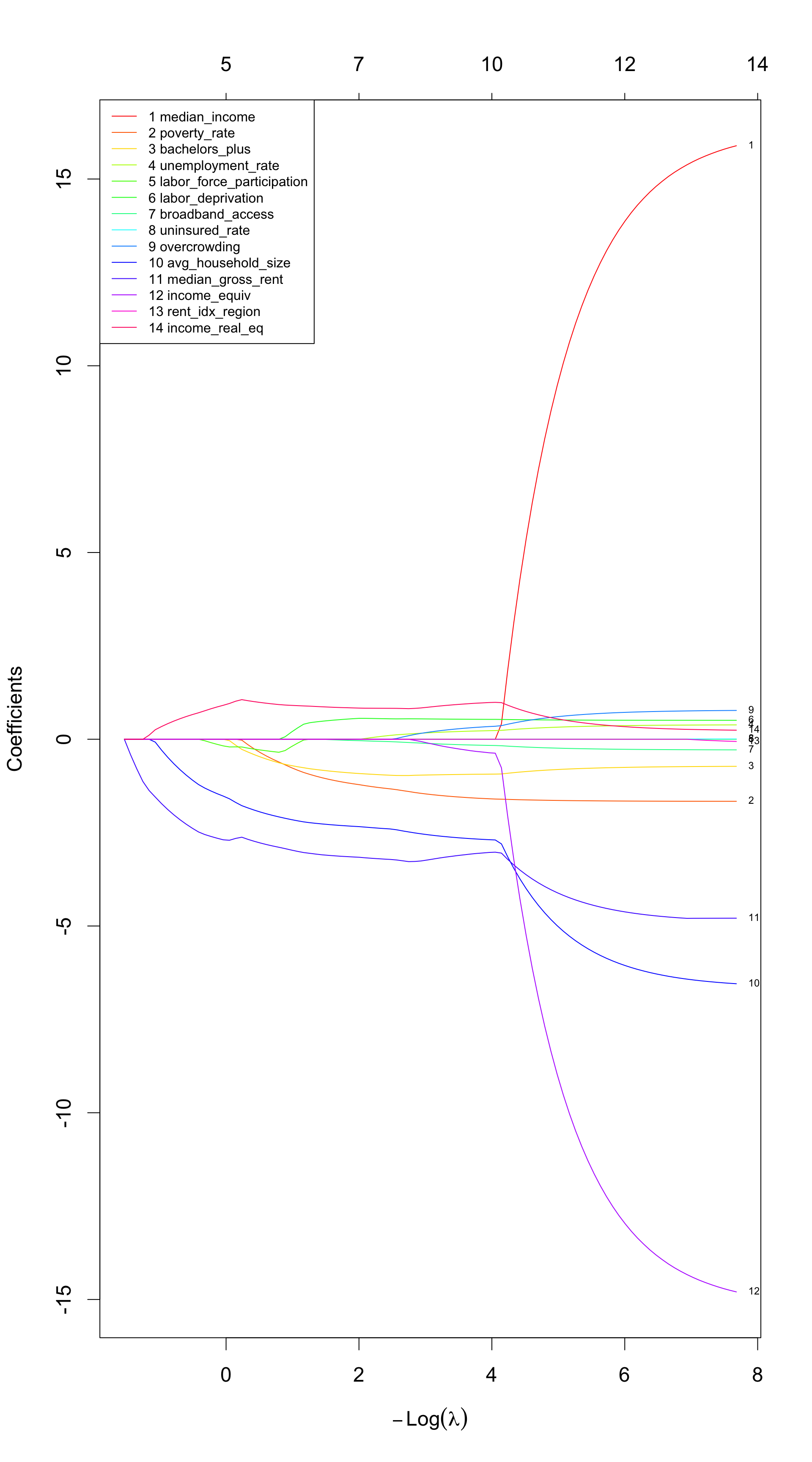}
    \caption{LASSO Coefficient Plot}
    \label{fig:LASSO-coeff-plot}
\end{figure}

After making the graphs, we decided to look into making a linear model for predicting suicide rate, but by only using economic variables as the predictors. Fitting the model, we got an $R^2$ of 0.501. This model, however, has much concern regarding multicollinearity, as some of the coefficients have VIFs of over 350. If we are purely concerned for prediction, then multicollinearity is not as much of a concern to us, assuming we are predicting in the region of the observed data (\cite{Kutner2005}). However, assuming multicollinearity is a concern, we choose LASSO with cross-validation (CV), as it can help mitigate instability due to correlated predictors and identify more important variables in the model for prediction (\cite{Tibshirani1996}). To perform LASSO here, we first standardized the predictors and then obtained the LASSO coefficient plot for this model (Figure \ref{fig:LASSO-coeff-plot}). We also obtained the best LASSO model (in terms of CV), which yielded the coefficients shown in Table \ref{tab:suicide-coeff-sizes-lasso}.

In Figure \ref{fig:LASSO-coeff-plot}, we see that as $\lambda$ increases, the variables that are zeroed out at the latest include Median Gross Rent, Real Equivalized Income, and Average Household Size.

\begin{table}[H]
    \centering
    \begin{tabular}{|c|c|}
    \hline
        Economic variable & Coefficient \\
        \hline
        Median Income & 0 \\
        \hline
        Poverty Rate & -1.07 \\
        \hline
        Bachelors Plus & -0.841 \\
        \hline
        Unemployment Rate & 0 \\
        \hline
        Labor Force Participation & 0 \\
        \hline
        Labor Deprivation & 0.500 \\
        \hline
        Broadband Access & -0.007 \\
        \hline
        Uninsured Rate & 0 \\ 
        \hline
        Overcrowding & 0 \\
        \hline
        Avg. Household Size & -2.28 \\
        \hline
        Median Gross Rent & 0 \\
        \hline
        Equivalized Income & 0 \\
        \hline
        Rent Index Region & -3.12 \\ 
        \hline
        Real Equivalized Income & 0.856 \\
        \hline
    \end{tabular}
    \caption{Economic variables' coefficient values for the model with suicide rate as response with the best in terms of CV}
    \label{tab:suicide-coeff-sizes-lasso}
\end{table}
After fitting the model, we gathered $R^2$ values for predicting each destructive variable, but only using economic variables as predictors, as we see in Table \ref{tab:r-squareds-for-destr-var-using-econ}. To note, all the $R^2$ values given are from estimated linear models built using ordinary least squares (OLS) and do not adjust for multicollinearity, unlike the model(s) associated with Figure \ref{fig:LASSO-coeff-plot}.

\begin{table}[H]
    \centering
    \begin{tabular}{|c|c|}
    \hline
        Response & $R^2$ \\
        \hline
        Suicide Rate & 0.501 \\
        \hline
        Drug Overdose Rate & 0.133 \\
        \hline
        Excessive Drinking & 0.524 \\
        \hline
        Insufficient Sleep & 0.452 \\
        \hline
        Smoking Rate & 0.676 \\
        \hline
        Injury Death Rate & 0.467 \\
        \hline
        Homicide Rate & 0.379 \\
        \hline
        Motor Vehicle Crash Death Rate & 0.429 \\
        \hline
        Firearm Fatality Rate & 0.504 \\
        \hline
        Physical Inactivity & 0.723 \\
        \hline
        Frequent Mental Distress & 0.702 \\
        \hline
    \end{tabular}
    \caption{$R^2$ table for each destructive variable as response, only using economic variables as predictors}
    \label{tab:r-squareds-for-destr-var-using-econ}
\end{table}

The highest $R^2$ values that were observed were when the response variable was Physical  Inactivity (0.723), Frequent Mental Distress (0.702), and Smoking Rate (0.676), respectively. In essence, those three health variables yielded the best linear fits when used as the response, while predictors were economic factors.
\section{Discussion and Conclusions}
\label{Discussion}

The PCA and clustering results suggest that some counties group by broader state-level economic patterns. However, interpreting these clusters politically would require adding explicit electoral or political variables and validating the classification.

Many negative correlations between economic variables and destructive variables are seen in the correlation analysis. For instance, in Figure \ref{fig:corr-matrix} and \ref{fig:destr-econ-graph}, we see the suicide rate being negatively correlated with Median Income, Labor Force Participation, Average Household Size, etc. In our LASSO linear model, we also see a negative coefficient for Bachelor Plus and Rent Index Region, and a positive coefficient for Labor Deprivation. Results in \cite{Naher_Rummel-Kluge_Hegerl_2020} are similar to these results, as the paper suggests an inverse relation between Socioeconomic Status and Suicide Rate.

With that being said, in our LASSO model, we do see a positive coefficient for Real Equivalized Income and a Negative Coefficient for Poverty Rate (somewhat contradicting what we said above). The sign for the Real Equivalized Income can be due to its formula; Equivalized Income is divided by the Rent Index Region. Since Equivalized Income is adjusted by the rent, high-suicide counties may have low income, but even lower rent, making for a prominent increase from Equivalized Income to Real Equivalized Income in the county. 

The negative coefficient for Poverty Rate could be due to other variables that are used in the Poverty Rate formula that are not in our model or are zeroed out. For instance, \cite{DAVICO2025116} showed a statistically significant association that populations of lower density tend to have higher suicide rates. Overcrowding in our data has a -0.42 correlation with Suicide Rate, but Overcrowding can be considered to negatively affect a county or population's poverty rate. Thus, in some instances, an increase in the poverty rate may be associated with a decrease in the suicide rate, since a variable used that derives poverty rate may increase poverty rate when it increases, but its increase is associated with a lowering of suicide rate.

In using LASSO, we were able to see, through its use of variable selection, how variables were chosen as shrinkage increased. This lets us get insight into what's redundant in our model and also lets us know what parameters are more useful in giving us good predictions. In our case, knowing what variables are more useful for predicting suicide rate has many positive gains. These variables may help identify economic conditions associated with higher suicide rates, but the results should not be interpreted as causal without further study.

Overall, the analysis suggests that county-level economic indicators are meaningfully associated with several adverse health outcomes in the western United States. The strongest linear relationships were observed for physical inactivity, frequent mental distress, and smoking rate, while suicide rate showed a moderate relationship with economic predictors. These results support the value of using county-level economic data for exploratory public health analysis, while also highlighting the need for caution because the models are observational, ecological, and affected by missing-data imputation and multicollinearity.

\section{Limitations} \label{sec:limitations}

\subsection{Imputation Drawbacks}
As previously mentioned, with our implementation of Regression Imputation (RI), it is to be noted that some results may be inaccurate and should not be treated as definitive. How the data can be inaccurate is further explained below.

\begin{itemize}
    \item For one, replacing the missing entries with column medians as we did, or alternatively using column means, modes, etc. for replacement at the start is a necessary step in this imputation process for our purposes of making a fully complete data set, but it does lead us to have more possible ways that the imputed data at the end can be inaccurate (since at the start, we are inputting counties that use column median values instead of regression predicted values for missing entries for fitting the models, which can very possibly be inaccurate).
    \item  Furthermore, with RI, counties or boroughs with barely any data to begin with before RI (such as Aleutians West Census Area in Alaska) result in the data from RI for such regions to have a high possibility of being inaccurate. This is obviously because the regression models may not have enough original information of such counties or boroughs for reliable imputation, and only mainly has medians to go off of (which can, again, be very inaccurate statistics from the real missing truth of the county).
\end{itemize}
With this being said, it is highly recommended to take the results and data not as definitive and not to make important decisions based on them. The results and data should be used to support further research in the area of study (or other areas) and for investigational purposes only.

\subsection{PCA and Clustering}
PCA was used as a way to get a general sense of the economic structure that any one county has in relation to other counties. This seemed to work well, as patterns in the PCA plot were seen. However, this method, if taken too seriously or literally, can cause issues and oversimplify the complex data at hand. 

Furthermore, when making clusters for the PCA data, possible better and more standard methods could've been used for better prediction. The idea for clustering here was to have a better visual understanding of the data, and shouldn't be extended to having competitive performance in prediction or classification.

\subsection{Linear Model Suitability}

Some variables in the linear models made may have been redundant to put in in the first place, such as having Equivalized Income, Average Household Size, and Median Income in the same model. This redundancy can be seen in the Figure \ref{fig:LASSO-coeff-plot}, as the coefficient paths for both equivalized income and median income, when ignoring the signs, are almost exactly the same. Future research should either remove such redundancy manually, by doing variable selection (such as using LASSO), or use other methods to make a more appropriate model.

What's also to note is that our $R^2$ values seen in Table \ref{tab:r-squareds-for-destr-var-using-econ} can only give us part of the picture, as all we did was do a basic linear model that didn't change at all when switching between different Destructive Behaviors as the response. There could be a more appropriate non-linear model or modified linear model for some of the Destructive Behaviors that gives a better fit.

\subsection{Coding}
To also be aware of, coding mistakes in the data gathering, imputation, and analysis are possible.  As such, awareness of these facts is encouraged through looking at the results and data.

\section{Disclosure Statement}

GenAI, such as Microsoft Copilot or OpenAI's ChatGPT 5.5 model, was used throughout the coding process for assistance with syntax, graphing, etc.

\section{Data Availability Statement}
\label{Data Availability}

Check the website https://noahdpj03.shinyapps.io/Web-Econ-disp/, where you can make your own choropleths by selecting a variable from the dataset and selecting the states wanted for the choropleth. Through this website, you can also download the data that was used for this project (either imputed or non-imputed) in the \textbf{Data} section.

\newpage

\bibliographystyle{apalike}
\bibliography{references}

\end{document}